\documentclass[preprint,12pt,sort&compress]{elsarticle}
\makeatletter
  \long\def\pprintMaketitle{\clearpage
  \iflongmktitle\if@twocolumn\let\columnwidth=\textwidth\fi\fi
  \resetTitleCounters
  \def\baselinestretch{1}%
  \printFirstPageNotes
  \begin{center}%
 \thispagestyle{pprintTitle}%
   \def\baselinestretch{1}%
    \Large\@title\par\vskip18pt
    \normalsize\elsauthors\par\vskip10pt
    \footnotesize\itshape\elsaddress\par\vskip10pt
    \end{center}%
  \gdef\thefootnote{\arabic{footnote}}%
  }
\makeatother
\usepackage{siunitx}
\usepackage{bm}
\usepackage{subcaption}
\usepackage{multirow}
\usepackage{booktabs}
\usepackage{lineno}
\usepackage{verbatim}
\usepackage{dcolumn}
\usepackage{longtable}
\usepackage{graphicx}
\usepackage{hyperref}
\usepackage{amsmath}
\usepackage{listings}
\usepackage{tikz}
\hypersetup{
	colorlinks=true,
	allcolors=blue
}
\newcommand*{\dif }{\mathop{}\!\mathrm{d}}
\newcommand{\jpsi}{J/\psi}
\newcommand{\piz}{\pi^{0}}
\newcommand{\ee}{e^{+}e^{-}}
\newcommand{\eexi}{\jpsi\to\Xi^{0}(\to\Lambda\gamma)\bar{\Xi}^0(\to\bar{\Lambda} \piz)}
\newcommand{\eebkg}{\jpsi\to\Xi^{0}\bar{\Xi}^0\to  \Lambda \piz+  \bar{\Lambda} \piz}
\newcommand{\XiRadDec} {\Xi^0\to \Lambda\gamma }
\newcommand{\XibRadDec}{\bar{\Xi}^0\to \bar{\Lambda}\gamma}
\newcommand{\XiHadDec} {\Xi^0\to \Lambda\piz}
\newcommand{\XibHadDec}{\bar{\Xi}^0\to \bar{\Lambda}\piz}
\newcommand{\LDec}{\Lambda\to p\pi^-}
\newcommand{\LbDec}{\bar{\Lambda}\to \bar{p}\pi^+}
\newcommand{\xiz}{\Xi^0}
\newcommand{\xib}{\bar{\Xi}^0}
\newcommand{\jpsiXX}{\jpsi\to\Xi^0\bar{\Xi}^0}
\newcommand{\mrec}{M_{\rm rec}}

\renewcommand{\thefootnote}{\fnsymbol{footnote}}

\DeclareSIUnit\clight{\text{\ensuremath{c}}}

\journal{Science Bulletin}

\begin{document}
\begin{frontmatter}
\title{\boldmath Measurement of the Decay $\ensuremath{\Xi}^{0}\ensuremath{\rightarrow}\ensuremath{\Lambda}\ensuremath{\gamma}$ with Entangled
	${\ensuremath{\Xi}}^0\bar{\ensuremath{\Xi}}^{0}$ Pairs}
	\author{BESIII Collaboration\footnote{Corresponding author (email: besiii-publications@ihep.ac.cn)}\footnote{Authors are listed at the end of this paper.}}
\end{frontmatter}
%\linenumbers

The ground-state hyperons have played a crucial role in the understanding of weak interactions. However, numerous unanswered questions persist regarding the weak decay mechanisms and the role of hyperons in
non-perturbative quantum chromodynamics (QCD). Specifically, weak radiative hyperon decays (WRHDs) provide valuable insights into the nature of
nonleptonic weak interactions. 
In general, the
radiative decays of a spin-$\frac{1}{2} $ hyperon  
consist of a parity conserving (P-wave) and a parity
violating (S-wave) amplitude. The decay
asymmetry parameter $\alpha_\gamma$, determined by the phase difference between S- and P-waves, can be measured from the asymmetric decay angle distribution of the
final state baryon and presents a longstanding puzzle in WRHDs~\cite{PhysRevLett.12.378}. Although decay asymmetries of $\Sigma^+\to p\gamma$ and $\Xi^{-}\to \Sigma^-\gamma$ decays
are expected to be zero under SU(3)
symmetry, experiments have measured non-zero values for these~\cite{PhysRevLett.14.154}. Various phenomenological models have been proposed to explain the
experimental
results~\cite{Niu_2020},
however, none of them provide a unified description of all WRHDs. To make further progress towards solving this puzzle, new and more
precise measurements of any WRHD are desired~\cite{Shi:2022dhw}.
On the other hand, the current observed combined charge conjugation and parity symmetries
($CP$) asymmetries are insufficient in explaining the observed matter-antimatter asymmetry.
 This leads to a surge in studies of $CP$ violation beyond weak hadronic decays. 
Weak radiative decays of $s$, $c$ and $b$ quarks are sensitive to new physics beyond the
Standard Model (SM) and may exhibit over
\SI{10}{\percent} $CP$ asymmetry in some models~\cite{Atwood:1997zr}. The large production rate of WRHDs in charmonium decay makes them experimentally attractive for these
studies.

The decay $\XiRadDec$ is a fundamental process for the study of WRHDs. 
Its $\alpha_\gamma$ serves as a crucial value to confirm the sources of the S-wave amplitude in WRHDs. 
On the other hand, a chiral perturbation theory (ChPT) for baryons can be constructed with only a few
input parameters~\cite{Shi:2022dhw}. In baryon ChPT, a low energy constant $C_\rho$ describes the direct photon emission contribution to the real part of the S-wave amplitude at tree level. To date, $C_\rho$ can only be
determined through the processes $\XiRadDec$ and $\Xi^{0}\to\Sigma^{0}\gamma$. Therefore, robust and precise experimental results are urgently
needed for improving the theory. Although studies of $\XiRadDec$ have been performed with
ever-increasing statistics at fixed target experiments~\cite{ParticleDataGroup:2024cfk}, all the measured branching fractions (BFs) have been reported as relative values to the BF of $\XiHadDec$. Furthermore, the systematic uncertainty is dominating the precision of BF and $\alpha_\gamma$ measurements.

The observation of hyperon transverse polarization in the decays $\jpsi\to B\bar{B} ~ (B=\Lambda,\Sigma^{\pm,0},\Xi^{-,0})$ at BESIII
has opened a new territory to study the character of hyperon
decays~\cite{BESIII:2018cnd}. The unique spin entanglement
information present in hyperon pairs enhances the statistical sensitivity to $\alpha_\gamma$ compared to fixed target experiments by several
times with the same sample sizes. Additionally, the application of the double-tag method
enables the measurement of absolute BFs with lower systematic uncertainty. These advantages have been validated in the studies of
the decays $\Lambda \rightarrow n\gamma$~\cite{BESIII:2022rgl} and
$\Sigma^+ \rightarrow p\gamma$~\cite{BESIII:2023fhs} at BESIII, where significant deviations of BFs from the world average
values~\cite{ParticleDataGroup:2024cfk} have been observed.

In this Letter, using the $\eexi$ decay based on \num{10087 \pm 44
	e6} $\jpsi$ events~\cite{BESIII:2021cxx} collected at \mbox{BESIII},
we report measurements of the absolute BF and the decay asymmetry parameter of
$\XiRadDec$, as well as a potential $CP$ asymmetry in the decay.
Throughout this Letter, charge conjugation (c.c.) is always
implied and $\Lambda$ is reconstructed using its decay $\Lambda\to p\pi^{-}$, unless explicitly noted otherwise. 

A detailed description of the \mbox{BESIII} detector can be found in the supplemental material. A double-tag method is employed for event selection. To search for the decay $\XiRadDec$, events of $\jpsiXX$
are tagged by reconstructing a $\xib$ signal (named `single-tag (ST)' events) using its dominant decay channel $\XibHadDec$. Subsequently, the signal $\XiRadDec$
is searched for in the system recoiling against the ST $\xib$, and the selected events are named `DT'. The absolute BF is calculated by $\mathrm{BF}(\XiRadDec) =
	\frac{N^{\mathrm{obs}}_{\mathrm{DT}}~\varepsilon_{\mathrm{ST}}}{N^{\mathrm{obs}}_{\mathrm{ST}}~\varepsilon_{\mathrm{DT}}} \frac{1}{\mathrm{BF}_{\LDec}}$ ,
where $N^{\mathrm{obs}}_{\mathrm{ST(DT)}}$ and $\varepsilon_{\mathrm{ST(DT)}}$ are the ST (DT) yields and the corresponding detection efficiencies
and $\mathrm{BF}_{\LDec}$ is the BF of $\LDec$.

A general formula for the differential cross sections of the signal $\eexi$ and the dominant background $\eebkg$ is detailed in the supplemental
material.
The decay parameters used in the formula include: 
$\alpha_\psi$ and $\Delta\Phi_{\psi}$ for  $\jpsi\to\xiz\xib$,
$\alpha_0$ ($\bar{\alpha}_0$) and $\Delta\phi_0$ ($\Delta\bar{\phi}_0$) for $\xiz\to\Lambda\piz$
($\xib\to\bar{\Lambda}\piz$), $\alpha_\gamma$ ($\bar{\alpha}_\gamma$) for $\XiRadDec$ ($\XibRadDec$), and $\alpha_{\Lambda}$
($\bar{\alpha}_{\Lambda}$) for $\LDec$ ($\LbDec$).
A  Monte Carlo (MC) simulated sample of generic $\jpsi$ decays is used to study the potential background.
The signal MC samples for ST signal $\ee\to\jpsi\to\Xi^{0}(\to
	\mathrm{anything})\xib(\to\bar{\Lambda}\pi^0)$, DT signal $\eexi$ and the dominant background $\eebkg$ are generated with the amplitude introduced above. The input decay parameters for the simulation are fixed to those of the latest measurement from
BESIII~\cite{BESIII:2023drj}, except for $\alpha_\gamma$ and $\bar{\alpha}_\gamma$ which are determined in this analysis.
The slight difference in reconstruction efficiencies between the signal MC samples and data is corrected with control
samples. Specifically, the selection efficiency for protons is studied with
the control sample $\jpsi\to p\bar{p}\pi^{+}\pi^{-}$, the one for $\pi^{-}$ and $\Lambda$ reconstruction with the control sample
$\jpsi\to\Xi^{-}(\to\Lambda\pi^-)\bar{\Xi}^+(\to\bar{\Lambda}\pi^+)$, the one for $\piz$ with the
control sample $\jpsi\to\xiz(\to\Lambda\piz)\xib(\to\bar{\Lambda}\piz)$, and the one for photon with the control sample $\jpsi\to\gamma\mu^{+}\mu^{-}$.
The correction coefficients are extracted as a function of the particle momentum and polar angle, and the efficiency-corrected signal MC
sample will be used to evaluate the efficiency and measure $\alpha_\gamma$.

Charged tracks and photon candidates are detected in the main drift chamber (MDC) and electromagnetic calorimeter (EMC), respectively, and are selected with the same requirements as in Ref.~\cite{BESIII:2023drj}.
Particle identification (PID) by combining the information of $\dif E/\dif x$ from MDC  and time-of-flight (TOF) from TOF detectors is applied to select
(anti-)proton candidates~\cite{BESIII:2023drj}, 
while charged tracks other than proton candidates are regarded as pions.
The $\bar{\Lambda}$ candidate is reconstructed with $\bar{p}\pi^+$ combinations, which are constrained to originate from a common vertex and are required to
have an invariant mass with $\left|M_{\bar{p}\pi^+}-M_{\bar{\Lambda}}\right|< \SI{6}{MeV/\clight^2}$, where $M_{\bar{\Lambda}}$ is the nominal 
$\bar{\Lambda}$
mass~\cite{ParticleDataGroup:2024cfk}.
The $\pi^0$ candidates are reconstructed with pairs of photons whose
invariant mass is within the interval $\SI{115}{MeV/\clight^2}
< M_{\gamma\gamma} < \SI{150}{MeV/\clight^2}$, and the momenta are updated by a kinematic fit constraining $m_{\gamma\gamma}$ to the nominal $\piz$ 
	mass~\cite{ParticleDataGroup:2024cfk}.

The ST $\xib$ candidate is reconstructed with the $\bar{\Lambda}\piz$ combination whose  invariant mass
($M_{\bar{\Lambda}\piz}-M_{\bar{p}\pi^+}+M_{\bar{\Lambda}}$) is closest to the nominal $\xiz$
mass ($M_{\xiz}$)~\cite{ParticleDataGroup:2024cfk}, and is required to fall in
$\left|M_{\bar{\Lambda}\piz}-M_{\bar{p}\pi^+}+M_{\bar{\Lambda}}-M_{\xiz}\right|< \SI{12}{MeV/\clight^2}$. The ST yield is extracted by performing a binned
maximum likelihood fit on the distribution of the recoiling mass, defined as $\mrec=\sqrt{(E_{\mathrm{cms}}-E_{\bar{p}\pi^+}-E_{\piz})^2/c^4
	-(\mathbf{p}_{\bar{p}\pi^+}+\mathbf{p}_{\piz})^2/c^2 }$. In the formula, $E_{\mathrm{cms}}$ is the center-of-mass energy, $E_{\bar{p}\pi^+}$ ($\mathbf{p}_{\bar{p}\pi^+}$) and $E_{\piz}$ ($\mathbf{p}_{\piz}$) are the
energies
(momenta) of $\bar\Lambda$ and
$\piz$ in the $\jpsi$ rest frame, respectively. The model for mass fitting is constructed as follows. First, a signal sample, whose $\pi^0$ is matched in the MC truth by requiring the angle between the generated and reconstructed $\piz$ directions less than \SI{20}{\degree}, is obtained.
Then, the signal is described with the $\mrec$ distribution of this signal sample.
The backgrounds of $\jpsi\to\piz\Lambda\bar{\Sigma}^{0}$+c.c., $\jpsi\to\Sigma^{0*}\bar{\Sigma}^{0*}$, and  the combinatorial background of the signal are
all described by
the shapes of the corresponding MC simulated samples. 
Other backgrounds are described with a third-order polynomial function.
To compensate the resolution difference between data and MC simulation, the signal shape is convolved with a Gaussian function.
The fit curves are shown in the supplemental material. The ST yields and the detection efficiencies evaluated with the corresponding signal MC samples are summarized in Table~\ref{tab:br}.

The signal $\XiRadDec$ is selected with the remaining charged and neutral tracks recoiling against the ST $\xib$ candidates. Events with at least one
$\Lambda\to p\pi^-$ candidate and one photon are selected for further DT analysis. A five-constraint (5C) kinematic fit is performed under the hypothesis
$\jpsi\to\Lambda\bar{\Lambda}\piz\gamma$, imposing overall energy-momentum conservation and requiring the $\piz$ candidate to have its nominal mass~\cite{ParticleDataGroup:2024cfk}.
The $\chi^2$ of the 5C kinematic fit
($\chi^2_{\mathrm{5C}}$) is required to be less than \num{40}.
%In  case of multiple candidate combinations in one event, only the one with the minimum $\chi^2_{\mathrm{5C}}$ is kept.

After the above selection, MC studies indicate that the dominant sources of background events are the processes
$\jpsi\to\xiz(\to\Lambda\piz)\xib(\to\bar{\Lambda}\piz)$ and $\jpsi\to\Sigma^{0}(\Sigma^{0*})+X$, from here on referred to as BKG-I and BKG-II, respectively, where $X$ denotes any possible particle. In particular, the decays $\jpsi\to\piz\Lambda\bar{\Sigma}^0$+c.c. and
$\jpsi\to\Sigma^{0*}\bar{\Sigma}^{0*}$ account for a large proportion of BKG-II.
A powerful discrimination of signal from BKG-II requires~\cite{BESIII:2023fhs} the distance $L$ between the event primary vertex and  the intersection point of $\Lambda$ and
$\bar{\Lambda}$ momentum vector to satisfy $L/\sigma_L>2$, where
$\sigma_L$ is the resolution of $L$. 
%This requirement rejects  \SI{90}{\percent} of BKG-II events with a
%signal loss of less than \SI{20}{\percent}.
The BKG-II events are further suppressed by the requirement of $\left | M _{\gamma\bar{\Lambda}}-M_{\bar{\Sigma}^0}\right|>\SI{12}{MeV/\clight^{2}}$, where
$M_{\gamma\bar{\Lambda}}$ is the invariant mass of $\gamma$ from the signal side and $\bar{\Lambda}$ from the ST side, and $M_{\bar{\Sigma}^0}$ is the nominal $\bar{\Sigma}^0$ mass~\cite{ParticleDataGroup:2024cfk}. 

The momentum of $\Lambda$ in the rest frame of $\xiz$ ($p_\Lambda$)  will
be used to determine the signal yield~\cite{BESIII:2022rgl,BESIII:2023fhs}.
Unbinned extended maximum likelihood fits are performed on the $p_\Lambda$ distributions, to determine the  DT yields. In the fit, the shapes of the signal channel
$\XiRadDec$  and of the background channel $\XiHadDec$   are described with the corresponding MC simulated shapes convolved with Gaussian functions,
respectively, and the remaining background is described with a first-order polynomial function.
The fits are performed individually for the processes $\XiRadDec$ and $\XibRadDec$, and then a simultaneous fit between the two c.c. channels is also carried out by assuming the same BF between $\XiRadDec$ and $\XibRadDec$. As summarized in
Table~\ref{tab:br}, the results are well consistent
with each other.

The decay asymmetry parameter is measured for the $\XiRadDec$ and $\XibRadDec$ decay channels, which after applying $\SI{0.17}{GeV/\clight}< p_{{\Lambda}} (
	p_{\bar{\Lambda}})<\SI{0.19} {GeV/\clight}$ include \num{371} and \num{391} events, respectively, with a signal purity of $\sim\SI{77}{\percent}$. 
The joint
likelihood function ($\mathcal{L}$) is constructed according to the decay amplitude, which incorporates the set of observable and a set of
characterized decay parameters $H=(\alpha_{\psi}, \Delta\Phi_{\psi},
	\alpha_{\gamma}, \bar{\alpha}_{0}, \Delta\bar{\phi}_{0}, \alpha_{\Lambda}, \alpha_{\bar{\Lambda}})$. The effect of the detection efficiency on
$\mathcal{L}$ is evaluated by the normalization factor $\mathcal{N}$, and is calculated with the efficiency corrected signal MC sample by using the importance
sampling method. The target function for the fit is $S=-\ln\mathcal{L}+\ln\mathcal{L}_{\mathrm{bkg}}$, where the
contributions of background $\mathcal{L}_\mathrm{bkg}$ are estimated with the corresponding MC samples, including $\jpsi\to\xiz(\to\Lambda\piz)\xib(\to\bar{\Lambda}\piz)$,
 $\jpsi\to\piz\Lambda\bar{\Sigma}^0$+c.c. and
$\jpsi\to\Sigma^{0*}\bar{\Sigma}^{0*}$ MC samples. The background likelihoods are normalized to the fitted
yields of the
corresponding components in the data sample. The fits are performed for the $\XiRadDec$ and
$\XibRadDec$ decays individually. Furthermore, a simultaneous fit is also performed assuming the
same magnitude but opposite sign for the decay asymmetry parameters between the
c.c. channels. All fit results are shown in Table~\ref{tab:br}.

The total systematic uncertainties of BFs and $\alpha_{\gamma}$ are studied separately. 
The total systematic uncertainties are summed in quadrature and are shown in Table~\ref{tab:br}
for the individual and c.c. combined results.
The dominant source  of systematic uncertainties  for $\alpha_{\gamma}$ and $\bar{\alpha}_{\gamma}$ is associated with the continuum background model.  This uncertainty is evaluated using a data-driven method and suffers large fluctuation, leading to much different values for $\alpha_{\gamma}$ and $\bar{\alpha}_{\gamma}$, while
much smaller than the statistical uncertainty. Details about the systematic uncertainties are given in the supplemental material.

Based on the above results, a $CP$ asymmetry observable is constructed:
\begin{equation}
	\begin{aligned}
		 & A_{CP}=\frac{\alpha_\gamma+\bar{\alpha}_\gamma}{\alpha_\gamma-\bar{\alpha}_\gamma}= -0.120 \pm 0.084_{\rm stat.}\pm 0.029_{\rm syst.}, \\
	\end{aligned}
\end{equation}
where the systematic uncertainty of $A_{CP}$ only include the uncorrelated
uncertainties. The result is consistent with zero within uncertainties. The BF of the $\XiRadDec$ decay is invariant under the combined $CP$ transformation;
therefore, it's not analyzed in this context~\cite{Bigi:2021hxw}. 
\begin{table}[htpb]
	\caption{Individual and c.c. combined BFs and $\alpha_\gamma$ measurement results of $\XiRadDec$ and $\XibRadDec$. The
		first uncertainties are statistical and the second ones are systematic, if present.}
	\label{tab:br}
		\begin{tabular}{cD{,}{\pm}{-1}D{,}{\pm}{-1}}
			\toprule
			Channels                                    & \multicolumn{1}{c}{$\XiRadDec$}                  & \multicolumn{1}{c}{$\XibRadDec$}                \\
			\midrule
			$N_{\mathrm{ST}}^{\mathrm{obs}}$            & \num{1400541},\num{1989}                         & \num{1611216},\num{2111}                        \\
			$\varepsilon_{\mathrm{ST}}$ (\si{\percent}) & \num{17.61},\num{0.01}                           & \num{19.77},\num{0.01}                          \\
			$N_{\mathrm{DT}}^{\mathrm{obs}}$            & \num{308},\num{21}                               & \num{330},\num{25}                              \\
			$\varepsilon_{\mathrm{DT}}$ (\si{\percent}) & \num{4.49},\num{0.02}                            & \num{4.92},\num{0.02}                           \\
			Individual BF ($10^{-3}$)                   & \multicolumn{1}{c}{\num{1.348\pm0.090\pm0.054}}  & \multicolumn{1}{c}{\num{1.326\pm0.098\pm0.066}} \\
			Combined BF ($10^{-3}$)                     & \multicolumn{2}{c}{$1.347\pm 0.066\pm0.054$}                                                       \\
			\midrule
			Individual $\alpha_\gamma$ ($\bar{\alpha}_\gamma$)                  & \multicolumn{1}{c}{\num{-0.652\pm0.092\pm0.016}} & 
			\multicolumn{1}{c}{\num{0.830\pm0.080\pm0.044}}                                                                                                  \\
			Combined $\alpha_\gamma$                    & \multicolumn{2}{c}{\num{-0.741\pm0.062\pm0.019}}                                                   \\
			\bottomrule
		\end{tabular}
\end{table}

In summary, we have performed a study of the weak radiative hyperon decay $\XiRadDec$ using $\Xi^0\bar{\Xi}^0$ pairs produced in \num{10087\pm44e6} $\jpsi$ events collected with 
the \mbox{BESIII} detector. Benefiting from the spin entanglement between $\Xi^0$ and $\bar{\Xi}^0$ from $J/\psi$ decays, the absolute BF and decay asymmetry parameter $\alpha_\gamma$
are measured with high precision. The absolute BF of this decay is $ (\num[tight-spacing =
		true, parse-numbers=false]{1.347 \pm 0.066_{\rm stat.}\pm 0.054_{\rm
				syst.}})\times 10^{-3}$, and its decay asymmetry parameter $\alpha_\gamma$ is
$\num[tight-spacing = true, parse-numbers=false]{-0.741 \pm 0.062_{\rm
				stat.}\pm 0.019_{\rm syst.}}$.  
This work represents the first study of this decay at an electron-positron collider. 
\begin{figure*}[htpb]
	\centering 
	\begin{subfigure}[b]{0.55\textwidth}
	\begin{center}
	\includegraphics[width=\linewidth]{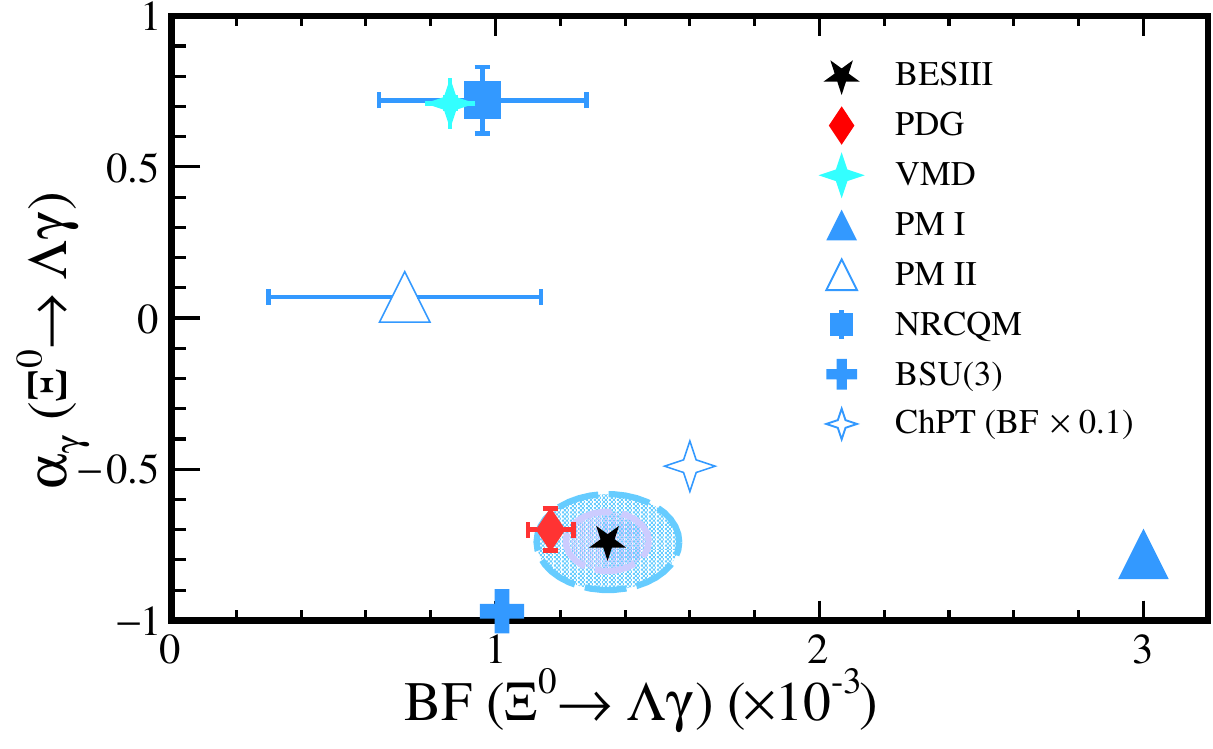}
	\caption{}
	\end{center}
	\end{subfigure}
	\begin{subfigure}[b]{0.4\textwidth}
	\begin{center}
	\includegraphics[width=\linewidth]{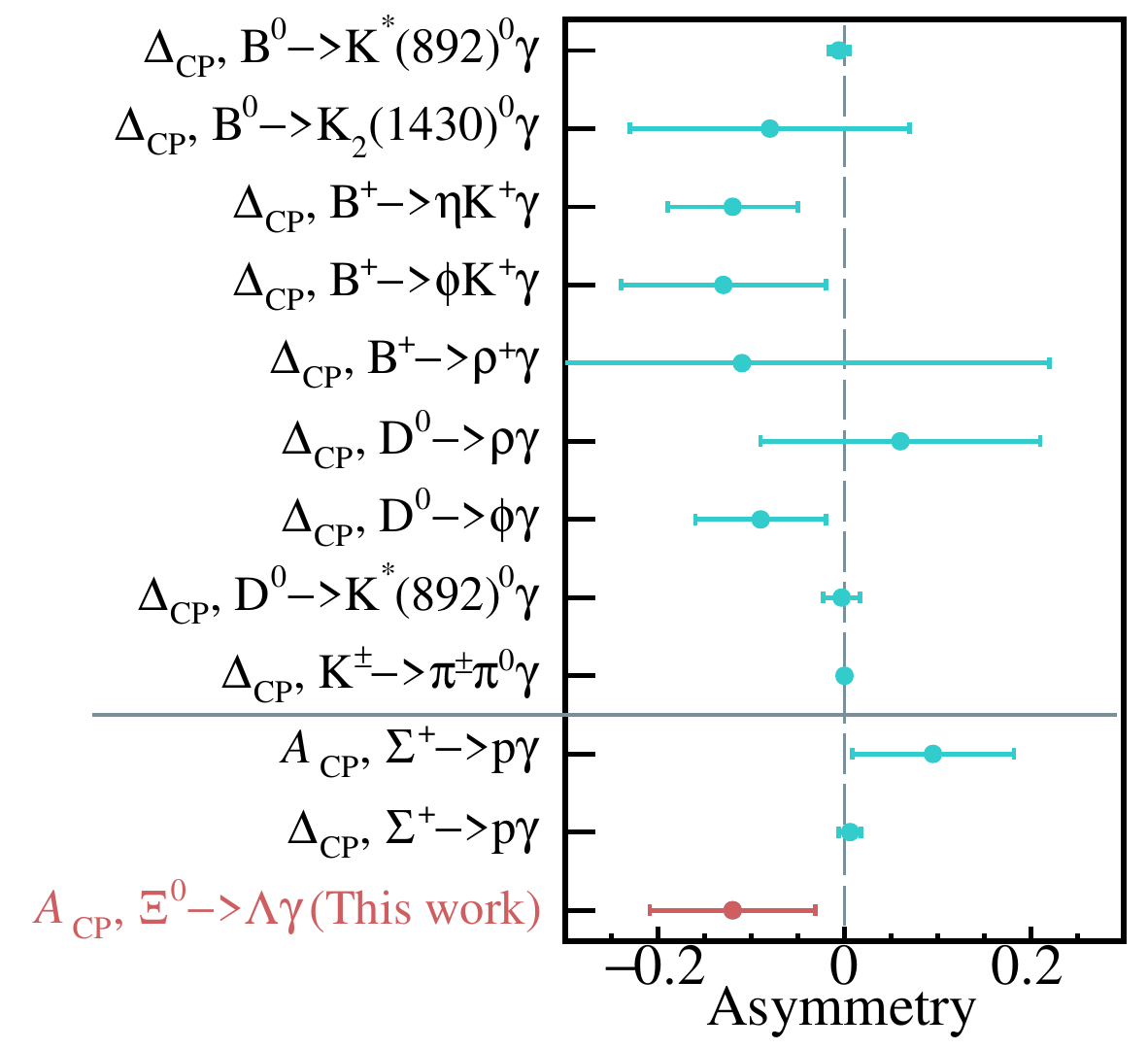}
	\caption{}
	\end{center}
	\end{subfigure}

\caption{(a) Distribution of $\alpha_{\gamma}$ versus BF of the
		$\XiRadDec$ decay. The black star denotes the results measured by
		this work and the blue contours correspond to the
		$\SI{68}{\percent}/\SI{95}{\percent}$ confidence-level of the
		results. The red diamond represents the PDG
		values~\cite{ParticleDataGroup:2024cfk} of the BF and
		$\alpha_\gamma$. Other symbols in blue or cyan~\cite{Niu_2020} stand for the results
		predicted by the vector meson dominance model
		(VMD)~\cite{zenczykowski_reanalysis_1991}, pole model~\cite{gavela_parity_1981,nardulli_pole_1987}, nonrelativistic constituent quark
		model (NRCQM)~\cite{Niu_2020}, broken SU(3) model
		(BSU(3))~\cite{Zenczykowski:2005cs} and
		a ChPT theory~\cite{borasoy_resonances_1999}. The BF result of ChPT theory is scaled down by a factor of 10 to fit
	within the figure. (b) Summary of the world average measurements of $CP$ asymmetry in weak radiative decays of mesons and baryons~\cite{ParticleDataGroup:2024cfk}.
	The asymmetry between the BFs (decay asymmetry parameters) of the particle and its antiparticle is denoted as $\Delta_{CP}$ ($A_{CP}$).
	The red dot represents the measurement result in this Letter.}
	\label{fig:result_com}
\end{figure*}

The results obtained for the BF and $\alpha_\gamma$ are consistent with the world average values~\cite{ParticleDataGroup:2024cfk} within $1.6\sigma$ and
$0.6\sigma$, respectively. The inconsistency of these results with various theory
predictions~\cite{zenczykowski_reanalysis_1991,gavela_parity_1981,nardulli_pole_1987,Niu_2020,Zenczykowski:2005cs,borasoy_resonances_1999} underscores the importance of refining the unified theory of weak hyperon
decay, which provides irreplaceable inputs to theories like baryon chiral perturbation theory~\cite{Shi:2022dhw} and forms the basis for research on physics beyond the
SM~\cite{Tandean:1999mg}
and $CP$ violation~\cite{He:2022bbs} in hyperon
decays.
Furthermore, this Letter presents the first search for $CP$ asymmetry in the decay $\XiRadDec$. While no evidence of $CP$ violation
is observed, the achieved sensitivity is at the same level to many measurements on weak radiative decays of mesons and is unique in baryon weak radiative
decays, as depicted in
Fig.~\ref{fig:result_com}. The potential for studying 
$CP$ asymmetry in an electron-positron collider is expected to be further explored at the proposed
Super Tau-Charm Facilities, with data samples enlarged by two orders of magnitude~\cite{Achasov:2023gey}. This
would undoubtedly open up new opportunities for investigating $CP$ violation in the baryon sector and advancing our understanding of fundamental physics.

\vspace{1cm}
\textit{Acknowledgments:} The authors thank Prof. L. S. Geng and Dr. R. X. Shi for their helpful discussion and constructive comments. The BESIII Collaboration thanks the staff of BEPCII and the IHEP computing center and the
	supercomputing center of USTC for their strong support. This work is supported in part by National Key R\&D Program of China under Contracts Nos.
	2023YFA1606000, 2023YFA1609400, 2020YFA0406300, 2020YFA0406400 ; National Natural Science Foundation of China (NSFC) under Contracts Nos. 11625523, 11635010,
	11735014, 11935015, 11935016, 11935018, 12025502, 12035009, 12035013, 12061131003, 12105276, 12122509,
	12192260, 12192261, 12192262, 12192263, 12192264, 12192265, 12221005, 12225509, 12235017, 12361141819; the Chinese Academy of Sciences (CAS) Large-Scale Scientific Facility Program; the CAS
	Center for Excellence in Particle Physics (CCEPP); Joint Large-Scale Scientific Facility Funds of the NSFC and CAS under Contract No. U1732263,
	U1832103, U1832207, U2032111; 100 Talents Program of CAS; CAS Key Research Program of Frontier Sciences under Contracts Nos. QYZDJ-SSW-SLH003,
	QYZDJ-SSW-SLH040; CAS Youth Team Program under Contract No. YSBR-101;  The Institute of Nuclear and Particle Physics (INPAC) and Shanghai Key Laboratory for Particle Physics and
	Cosmology; German Research Foundation DFG under Contracts Nos. 455635585, FOR5327, GRK 2149; Istituto Nazionale di Fisica Nucleare, Italy; Ministry of
	Development of Turkey under Contract No. DPT2006K-120470; National Research Foundation of Korea under Contract No. NRF-2022R1A2C1092335; National Science
	and Technology fund of Mongolia; National Science Research and Innovation Fund (NSRF) via the Program Management Unit for Human Resources \&
	Institutional Development, Research and Innovation of Thailand under Contract No. B16F640076; Polish National Science Centre under Contract No.
	2019/35/O/ST2/02907; The Swedish Research Council; U. S. Department of Energy under Contract No. DE-FG02-05ER41374.

%\bibliographystyle{test}
%\bibliography{bibitem.bib}

\begin{center}
	\textbf{BESIII Collaboration}
\end{center}
\begin{small}
		M.~Ablikim$^{1}$, M.~N.~Achasov$^{4,c}$, P.~Adlarson$^{76}$, O.~Afedulidis$^{3}$, X.~C.~Ai$^{81}$, R.~Aliberti$^{35}$, A.~Amoroso$^{75A,75C}$,
		Q.~An$^{72,58,a}$, Y.~Bai$^{57}$, O.~Bakina$^{36}$
% ...
\\

\noindent{\it
		$^{1}$ Institute of High Energy Physics, Beijing 100049, People's Republic of China\\
% ...
$^{3}$ Bochum  Ruhr-University, D-44780 Bochum, Germany\\
		$^{4}$ Budker Institute of Nuclear Physics SB RAS (BINP), Novosibirsk 630090, Russia\\
% ...
$^{35}$ Johannes Gutenberg University of Mainz, Johann-Joachim-Becher-Weg 45, D-55099 Mainz, Germany\\
		$^{36}$ Joint Institute for Nuclear Research, 141980 Dubna, Moscow region, Russia\\
% ...
$^{57}$ Southeast University, Nanjing 211100, People's Republic of China\\
		$^{58}$ State Key Laboratory of Particle Detection and Electronics, Beijing 100049, Hefei 230026, People's Republic of China\\
% ...
$^{72}$ University of Science and Technology of China, Hefei 230026, People's Republic of China\\
% ...
$^{75}$ University of Turin and INFN, (A)University of Turin, I-10125, Turin, Italy; (B)University of Eastern Piedmont, I-15121, Alessandria, Italy; (C)INFN, I-10125, Turin, Italy\\
		$^{76}$ Uppsala University, Box 516, SE-75120 Uppsala, Sweden\\
% ...
$^{81}$ Zhengzhou University, Zhengzhou 450001, People's Republic of China\\
	$^{a}$ Deceased\\
% ...
$^{c}$ Also at the Novosibirsk State University, Novosibirsk, 630090, Russia\\
% ...
}
\end{small}
%% ends here %%

\end{document}